# Digitally Programmable Photochromic Hydrogel Contact Lenses as Light-Adaptive Artificial Irises

Asad Nauman†, Tristen Moncada, Guli Gulinihali, Muhammad Waleed Khalid, Yeshaiahu Fainman, and Abdoulaye Ndao*

[1]*Department of Electrical and Computer Engineering, University of California, San Diego, La Jolla, CA 92093, USA*

* Abdoulaye Ndao: a1ndao@ucsd.edu

* Corresponding author. Tel.: +1(858)-534-2721.

**ABSTRACT**

Excessive exposure to ultraviolet (UV) radiation is associated with a range of ocular pathologies, motivating the development of soft optical devices that can dynamically regulate incident light. In the human eye, this adaptive optical functionality is performed by the iris, which modulates pupil size to control retinal irradiance in response to ambient illumination. Here we present a photochromic contact–lens–based artificial iris that mimics this biological light-adaptation mechanism through reversible, spatially programmable modulation of optical transmission with intrinsic UV blocking. Photochromic dyes are embedded within a biocompatible hydrogel matrix, while the cross-linked network is patterned using a digital micromirror device (DMD)– based grayscale UV lithography to encode controlled radial gradients in dye switching. This approach generates iris-like attenuation profiles that emulate pupil-dependent light regulation while enabling customizable iris geometries and transmission patterns. The resulting lenses exhibit rapid and reversible UV-induced darkening with position-dependent kinetics, enabling continuous modulation of transmitted light. The photoresponse remains stable over repeated activation cycles without measurable fatigue. The patterned lenses maintain mechanical stability, controlled swelling, and wettability suitable for contact lens applications. This platform combines programmable photochromism and hydrogel optics to enable light-adaptive lenses that mimic key functions of the human iris.

## 1. Introduction

The human iris provides an essential light-adaptive function by dynamically constricting and dilating the pupil to regulate the amount of light reaching the retina [1-3]. This self-regulating aperture control enhances visual acuity and reduces glare under fluctuating illumination [4-6]. Consequently, iris defects not only impair optical performance but also affect visual comfort and social interaction [8-10].

Soft artificial irises integrated with contact lenses offer a noninvasive strategy for managing congenital conditions such as aniridia, coloboma, and albinism, as well as trauma-induced iris loss [11-13]. Although surgically implantable artificial irises and cosmetic soft-lens solutions can restore appearance, they do not provide dynamic regulation of incident light [14,15]. More recently, soft and customized implants (HumanOptics, Inc.) aim to achieve better cosmetic integration, yet they lack light-adaptive pupil control [16, 17]. These limitations underscore the need for advanced artificial irises that can dynamically adjust to ambient light and modulate transmission in a manner that more closely replicates the natural iris.

To address the dynamic light modulation, recent pupil-scalable systems based on liquid-crystal elastomers or electrowetting require electrical actuation, raising safety concerns for continuous ocular wear [18-23]. Light-driven mechanical actuation has also been explored [24-30], but the deformation-based aperture change differs fundamentally from the rapid, reversible optical modulation performed by the natural iris.

Photochromic materials provide a more biomimetic alternative by enabling self-regulated modulation of light transmission without external power [31-34]. Commercial photochromic contact lenses (Johnson & Johnson Vision Care, Inc.) demonstrate the feasibility of this concept; yet most reported designs regulate light uniformly across the lens, lacking the spatially graded response necessary to emulate natural pupil dynamics [35]. This highlights substantial opportunities for advancing the development and application of photochromic contact lenses that mimic human iris-like functionality.

Conventional contact lenses are typically fabricated by spin casting, casting molding, or lathe cutting [36-38]; however, recent efforts have shifted toward additive manufacturing to enable greater design flexibility and patient-specific customization. Among available additive manufacturing approaches, vat photopolymerization has emerged as particularly suitable for contact lens production [39-41]. Researchers have demonstrated the fabrication of clear lenses as well as lenses incorporating color–vision–correcting dyes, achieving favorable optical transmission and hydration properties [42, 43]. However, the layer-by-layer material addition in 3D printing produced a staircase effect on the contact lens surface. Artifacts from layer-by-layer printing could be mitigated through post-processing, such as dip coating [44]. Building on this, a continuous liquid film–confined printing strategy has been suggested to further improve surface quality [45]. Additional studies have shown that 3D printing can localize dye deposition in customizable patterns within the lens matrix. However, existing additive manufacturing techniques struggle to control the switching speed of the photochromic dye positionally by selectively adjusting the cross-linking density of the hydrogel for contact lenses.

Here, we introduce photochromic artificial iris implemented in a soft hydrogel contact lens that achieves light-adaptive pupil-aperture control through spatially programmable and reversible molecular switching. Photochromic dyes are uniformly dispersed within a biocompatible hydrogel and selectively crosslinked using digital micromirror (DMD)–based UV patterning to generate a radial gradient in hydrogel cross-linking density, and consequently, the photochromic dye switching kinetics. This structured distribution enables iris-like attenuation profiles, dynamic modulation of effective aperture size, and robust UV blocking. Mechanical, swelling, optical, and cycling analyses confirm that the lenses maintain the material requirements of conventional hydrogels while exhibiting rapid, repeatable, and spatially resolved photoresponse. The fabrication method is compatible with multiple dye chemistries and can be readily integrated into cosmetic or therapeutic lenses, providing a versatile platform for both functional vision support and aesthetic.

## 2. Results and discussion

### *2.1. Conceptual illustration of artificial iris and its fabrication*

Figure 1A depicts the natural pupillary response, where the iris adjusts the aperture size in real time according to ambient illumination. Under low-light conditions (dark), the dilator pupillae muscles expand the pupil to roughly 4~8 mm, allowing more light to reach the retina. In contrast, under bright illumination, the sphincter pupillae muscles constrict the pupil to approximately 2~4 mm, reducing retinal light exposure. This reversible biological reflex optimizes visual acuity, mitigates glare, and protects ocular tissues from excessive radiation. The anatomy of the eye reveals the functional trajectory of incident light from the cornea through the pupil to the retina, emphasizing the significance of iris-mediated modulation in maintaining retinal health and visual clarity. To mimic this biological reflex action of the eye, photochromic dyes comprising of spiropyran compounds (colorless form), which transform to a merocyanine (colored) form upon UV exposure (Figure 1B (i)) are introduced to opto-chemically control the light transmission to retina. Figure 1B(ii) illustrates the radially patterned gradient cross-linking density of the hydrogel, enabling spatially controlled transition in photochromic dye switching efficiency to mimic the natural iris. In this design, the central region is engineered with a lower crosslinking density, while the peripheral region possesses progressively higher crosslinking density. Because photochromic molecular switching depends on available free volume and polymer-chain mobility, the lower-crosslinked central region allows stronger and faster coloration under UV exposure, whereas the more highly crosslinked outer region restricts molecular motion and therefore exhibits weaker switching.

This spatial difference in switching efficiency creates a radial optical gradient without requiring any mechanical deformation. Upon UV illumination, the central region darkens more strongly while the peripheral region remains relatively transparent, thereby reducing the effective aperture size in a manner analogous to natural pupil constriction. This produces an artificial, light-responsive iris capable of modulating the effective pupil size without mechanical deformation.

The cross-sectional scanning electron microscopic (SEM) image highlights the material architecture, where spatial heterogeneity in cross-linking density produces a gradient optical response. Areas with denser cross-linking exhibit more compact and tightly packed microdomains, which hinder dye isomerization and reduce switching efficiency, whereas regions with lower crosslinking density show larger and more open domains that provide sufficient space for photochromic molecular transformation. These morphological variations correlate directly with the observed optical gradients under UV exposure.

When the lens is exposed to incident UV (Figure 1B(iii)), the radially patterned photochromic layer develops a spatially varying opacity, which is strongest in the central region and progressively weaker toward the periphery. As a result, the contact lens exhibits a pupil-like darkened zone that constricts the effective aperture, analogous to the natural iris response. When the UV stimulus is removed, the colored state reversibly bleaches, causing the central region to recover higher transmission and effectively dilate the aperture. This dynamic and reversible modulation enables real-time light-adaptive regulation without electrical input or mechanical actuation.

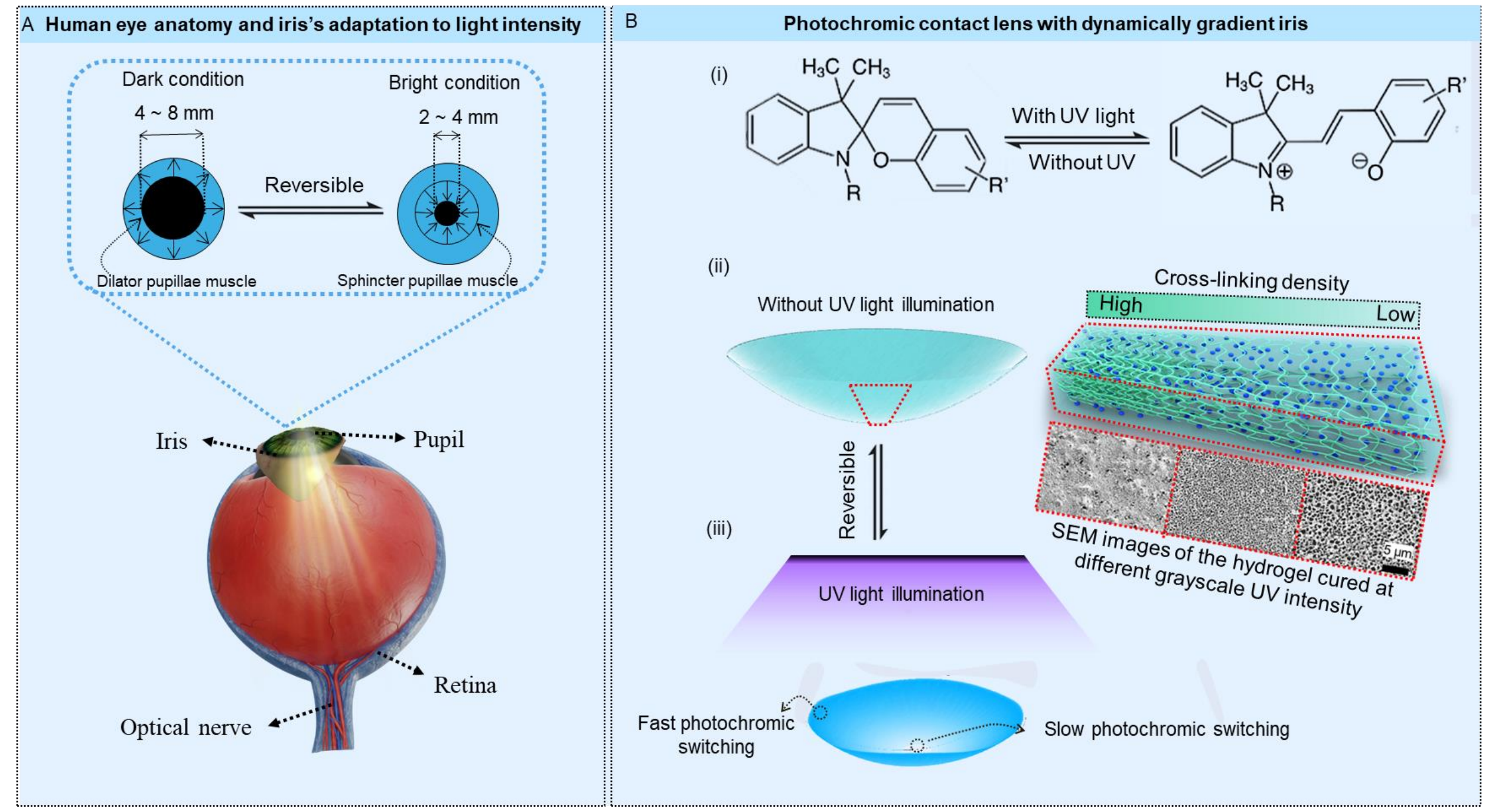

**Fig. 1.** Concept and design of a photochromic contact lens mimicking the adaptive human iris, (A) Schematic of human iris physiology, showing reversible pupillary constriction (2~4 mm) and dilation (4~8 mm) in response to light intensity, governed by sphincter and dilator muscles under different environmental lights, showing the reversible pupil dilation and contraction, (B) chemical structure of (i) photochromic dye in without UV-activated state (spiropyrane → SP, colorless) and in UV–activated state (merocyanine → MC, colored), (ii) contact lens in transparent colorless state without UV illumination, where the cross-linking density (shown in red dashed line) increases from the outer edge toward the center of the contact lens while photochromic dye distribution remain equal from edge to center. The SEM images corresponding to different cross-linking densities are shown. (iii) UV illumination induces spatially dependent darkening, modulating transmitted light analogously to the natural human iris.

Figure 2A shows the molecular components of the hydrogel precursor formulation, consisting primarily of 2-hydroxyethyl methacrylate (HEMA) as the hydrophilic monomer, poly(ethylene glycol) diacrylate (PEGDA) as the crosslinker, and dispersed photochromic dye molecules. HEMA provides softness, oxygen permeability, and hydration that are compatible with conventional contact lenses, while PEGDA establishes a tunable polymeric network whose crosslink density governs both mechanical integrity and swelling properties. The photoinitiator, diphenyl(2,4,6-trimethylbenzoyl)phosphine oxide (TPO), undergoes homolytic cleavage under UV illumination to generate free radicals, initiating chain-growth polymerization of the methacrylate groups and simultaneously immobilizing the photochromic dye within the hydrogel matrix. Consequently, generating a polymer network during UV curing, where the uniformly mixed monomers and dyes become covalently trapped in the crosslinked structure. The HEMA: PEGDA precursor ratio of 35:1 was selected to balance mechanical softness, water uptake, structural stability, and photochromic switching performance for contact lens applications. Previous studies commonly used higher PEGDA contents (e.g., 1:1 HEMA: PEGDA), which generate densely crosslinked hydrogels with higher modulus and lower swelling, making them less suitable for soft ophthalmic lenses [45]. By reducing the PEGDA content, the present formulation lowers crosslink density, resulting in improved hydration, lower stiffness, and greater polymer-chain mobility. This enhanced local free volume also facilitates reversible photochromic isomerization, leading to faster coloration/bleaching kinetics and stronger UV-responsive optical modulation. Therefore, the 35:1 ratio provides a more suitable hydrogel network for soft, light-adaptive contact lens applications. The cartoon depiction in Figure 2A highlights the molecular-level distribution of dye molecules, allowing for optical function while preserving the hydrogel's elasticity and transparency. Figure 2B shows the preparation of the precursor solution. Photochromic dyes are introduced into the monomer mixture and dispersed using a magnetic stirrer to ensure homogeneity prior to polymerization. Thorough dispersion minimizes aggregation of dye particles, which is critical for achieving uniform optical performance and preventing scattering artifacts. The resulting mixture forms a stable, photocurable solution ready for mold-based patterning. Figure 2C outlines the fabrication workflow. First, a transparent female mold (Figure 2C(i)) is prepared to define the anterior lens surface. The dye-containing hydrogel precursor is dispensed into the mold cavity, Figure 2C(ii), after which a transparent male mold, Figure 2C(iii), is positioned to create the desired lens curvature and thickness.

In Figure 2D, the mold assembly is subjected to patterned UV exposure using a digital micromirror device (DMD). By digitally controlling the projected UV intensity and spatial distribution, it becomes possible to engineer localized variations in polymerization kinetics and dye fixation. The complete optical setup of DMD is shown in Figure S1. This strategy directly enables the formation of radially graded regions within the hydrogel, which is essential for constructing an artificial iris architecture where optical density varies from the center toward the periphery. The DMD–based approach enables submillimeter precision, high reproducibility, and compatibility with various dye chemistries. Similar principles of spatial patterning underpin a broad range of engineered photonic elements, from planar dielectric lenses [49] and broadband meta-beam-splitters [50] to inverse-designed plasmonic nanoantennas [51], where locally varying geometry or material response encodes the desired optical function. Finally, the fully polymerized photochromic contact lens is demolded. The inset images demonstrate the clarity and structural integrity of the fabricated lens, along with visible patterned coloration when activated. The resulting devices retain the mechanical softness, optical transparency, and water content required for ocular comfort while exhibiting spatially tunable photochromic behavior suitable for artificial iris applications. Our fabrication design provides a platform that merges photo-initiated hydrogel chemistry, controlled cross-linking network and photo-switching of dye, and digitally patterned UV curing to produce advanced photo-responsive contact lenses (Figure 2E). The ability to encode graded or spatially selective photochromic regions within a soft hydrogel matrix represents a significant step toward the development of functional artificial irises and other adaptive optical devices for ophthalmic use.

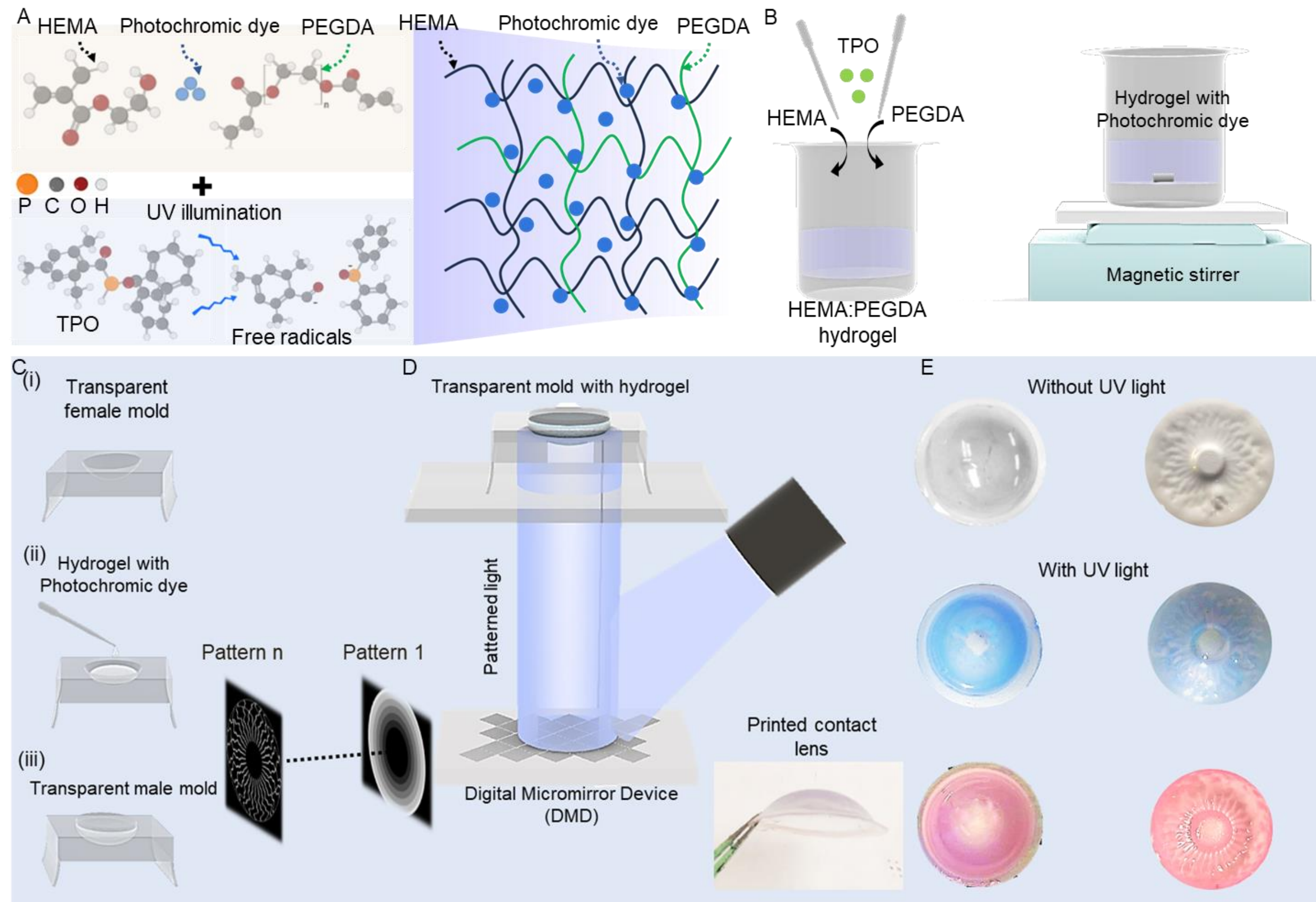

**Fig. 2.** Fabrication of soft iris contact lens, (A) chemical structures of HEMA, PEGDA, and the photochromic dye, along with the photoinitiator (TPO). Upon UV exposure, TPO generates free radicals that initiate polymerization, embedding photochromic dye molecules within the hydrogel network. (B) hydrogel precursor preparation: photochromic dye is dispersed into the monomer solution under magnetic stirring prior to UV curing. (C) realization of a gradiently cross-linked photochromic contact lens. (i) A transparent female mold is prepared. (ii) The photochromic dye-containing hydrogel precursor is dispensed into the mold. (iii) A transparent male mold is placed to define the lens geometry. (D) Patterned UV illumination via a digital micromirror device (DMD) selectively cures the hydrogel. The patterned photochromic contact lens is demolded, with representative fabricated lenses shown in the inset. (E) Fabricated contact lenses with customized gradient photochromic switching and patterns.

*2.2. Composition and Conformability of Artificial Iris*

To evaluate how the commercial photochromic powders integrate within the hydrogel, we first examined the microstructure and chemistry of the dyes and dyed contact lens. Figure 3A and B present the scanning electron microscope (SEM) images of the blue and pink photochromic dyes, respectively, showing that both materials consist predominantly of spherical microparticles with relatively uniform morphology. The high-magnification SEM images were acquired from representative regions of the same sample but do not correspond to the exact locations shown in the low-magnification images. After incorporation into the hydrogel lens matrix, the dyed contact lens surfaces (Figure 3D) exhibit dispersed particulate domains embedded within the polymer network, whereas the pristine clear contact lens (Figure 3C) shows a comparatively smooth and homogeneous surface. The absence of severe aggregation, cracking, or phase separation after dye incorporation indicates good compatibility between the photochromic dye and the HEMA/PEGDA matrix, while preserving the overall lens structure.

Fourier transform infrared (FTIR) spectroscopy further supports these morphological observations. The spectra of the isolated blue and pink dye powders display characteristic organic functional group absorptions, including broad bands in the 3200–3400 $cm^{-1}$ region assigned to O–H/N–H stretching vibrations, peaks near 3020–3060 $cm^{-1}$ corresponding to aromatic C–H stretching, and bands around 2920–2960 $cm^{-1}$ attributed to aliphatic C–H stretching vibrations. Strong absorptions in the 1600–1500 $cm^{-1}$ region are associated with aromatic C=C stretching and conjugated ring vibrations, while peaks in the 1250–1000 $cm^{-1}$ range arise from C–O, C–N, or ether-related stretching modes typical of the dye carrier matrix. Additional bands below 900 $cm^{-1}$ are attributed to aromatic C–H bending vibrations.

In contrast, both the clear contact lens and the dye-loaded contact lens exhibit nearly identical spectral profiles, demonstrating that the characteristic structure of the HEMA/PEGDA hydrogel network is maintained after dye incorporation. The lens spectra show the expected carbonyl (C=O) stretching band near ~1720 cm$^{-1}$ from ester groups of HEMA/PEGDA, C–O–C stretching vibrations in the 1100–1250 cm$^{-1}$ region, and broad O–H absorption associated with hydrated hydrogel domains. Only minor intensity variations are observed for the dyed lens, which are attributed to the low dye concentration rather than chemical modification of the polymer network. These results confirm that the photochromic dye is physically incorporated into the lens matrix without significantly altering the underlying hydrogel chemistry.

The mechanical response of the lenses was next characterized to verify that embedding photochromic dyes does not compromise softness or flexibility. Stress–strain curves in Figure 4A(i) show typical soft-hydrogel behavior for both undyed and dyed samples. The tensile modulus extracted at small strain (Figure 4A(ii)) remains within the range expected for commercial hydrogel contact lenses. These results indicate that the artificial iris retains the compliant, deformable character required to conform to the corneal surface and withstand routine handling.

Hydration and surface wettability measurements further confirm that the artificial iris lenses remain compatible with the ocular environment. Time-dependent swelling in both deionized water and PBS (Figure 4B (i)) shows a rapid initial increase followed by saturation, which is characteristic of diffusion-limited water transport in crosslinked hydrogels. The early stage is dominated by water diffusion into existing free volume, while the later stage reflects a balance between osmotic driving forces and elastic resistance of the polymer network. The equilibrium water absorption is essentially unchanged by dye incorporation, indicating that the dye does not alter network hydrophilicity, effective crosslink density, or mesh size. The slightly higher swelling in PBS compared to water is attributable to ionic osmotic effects, but its small magnitude suggests weak ion–polymer interactions, confirming preservation of the intrinsic hydration properties of the contact lens. The oxygen permeability of the photochromic contact lenses (Figure 4B (ii)) is lower than that of commercial lenses [46]. Commercial silicone hydrogel lenses exhibit particularly high oxygen permeability because silicone polymer chains provide an additional pathway for oxygen transport even at lower water contents. In contrast, conventional hydrogel lenses generally show lower oxygen permeability, where Dk is strongly dependent on water content. For example, the commercial hydrogel material etafilcon A, with a water content of 58%, has a reported Dk of 21.5, whereas polymacon, with a lower water content of 38%, exhibits a Dk of 8.5 [47]. The photocrhomic contact lenses developed in this study show an oxygen permeability of 12, which is higher than that of polymacon and lesser than etafilcon. Further improvement in oxygen permeability may be achieved by increasing the water content of the printed lenses or by incorporating silicone components to develop a silicone hydrogel-based system. Contact angle images in Figure 4C (i) show that all lens variants exhibit contact angles well below 90°, indicating hydrophilic surfaces. Although the addition of dye induces only a slight shift in contact angle Figure 4C (ii), the values lie within the range of commercial hydrogel contact lenses. For context, commercial contact lenses typically exhibit contact angles in the 44–79° range.

Overall, the combined structural, chemical, mechanical, and swelling data demonstrate that digitally patterned photochromic loading can be integrated into a soft hydrogel contact lens without sacrificing the core material properties needed for comfort and safe wear. The present device is based on HEMA-containing hydrogel materials widely used in contact lens and biomedical applications because of their favorable cytocompatibility, optical transparency, and ocular tolerance after proper curing. Previous studies have reported low cytotoxicity and good tissue compatibility for crosslinked HEMA hydrogels, suggesting that the developed lens is expected to be biocompatible under normal use conditions [48].

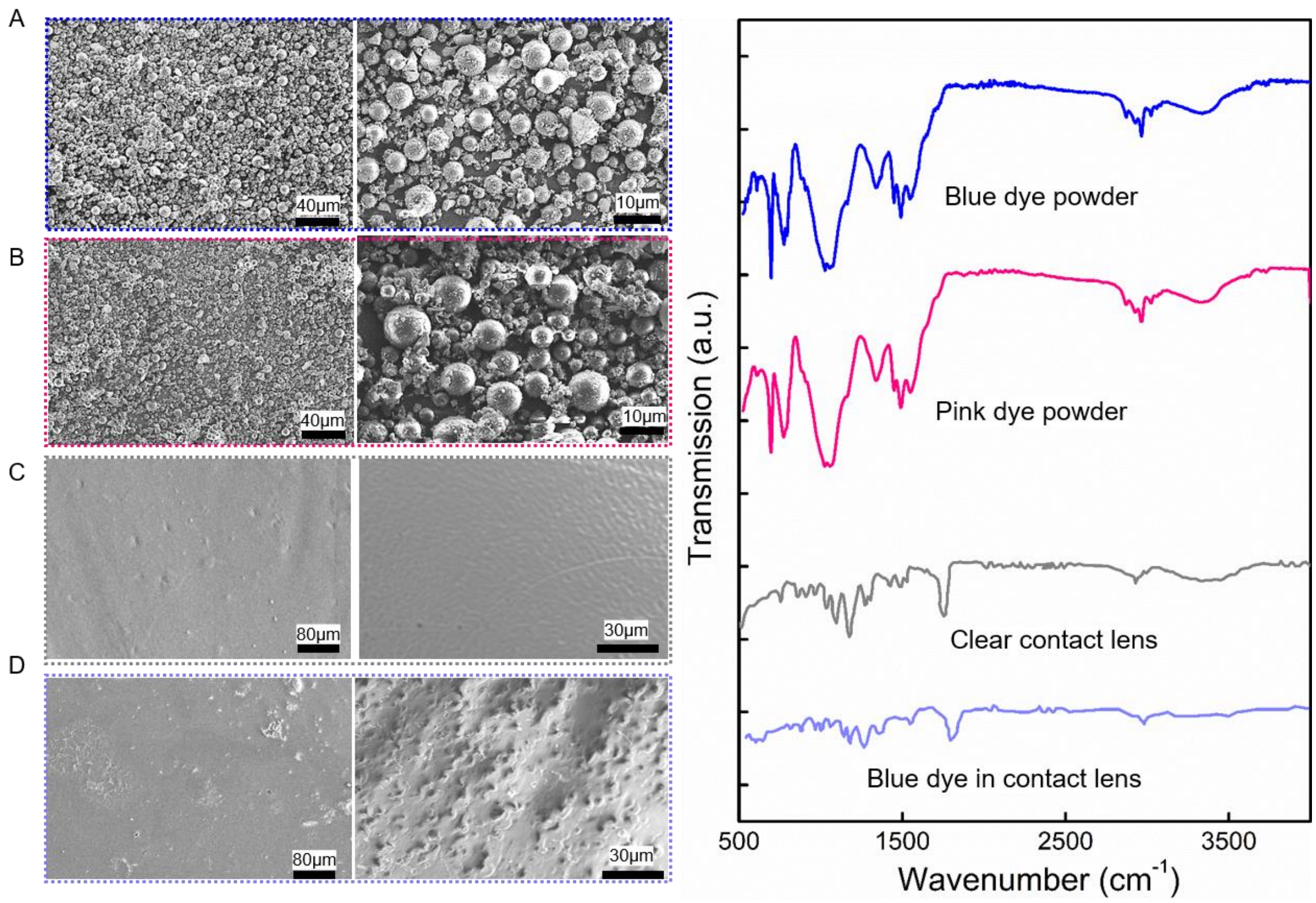

**Fig. 3**. Morphology and chemical characterization of photochromic dyes and contact lens. SEM images and corresponding FTIR of photochromic dye (A) blue dye powder, (B) pink dye powder, (C) pristine clear contact lens, (D) contact lens with embedded blue dye.

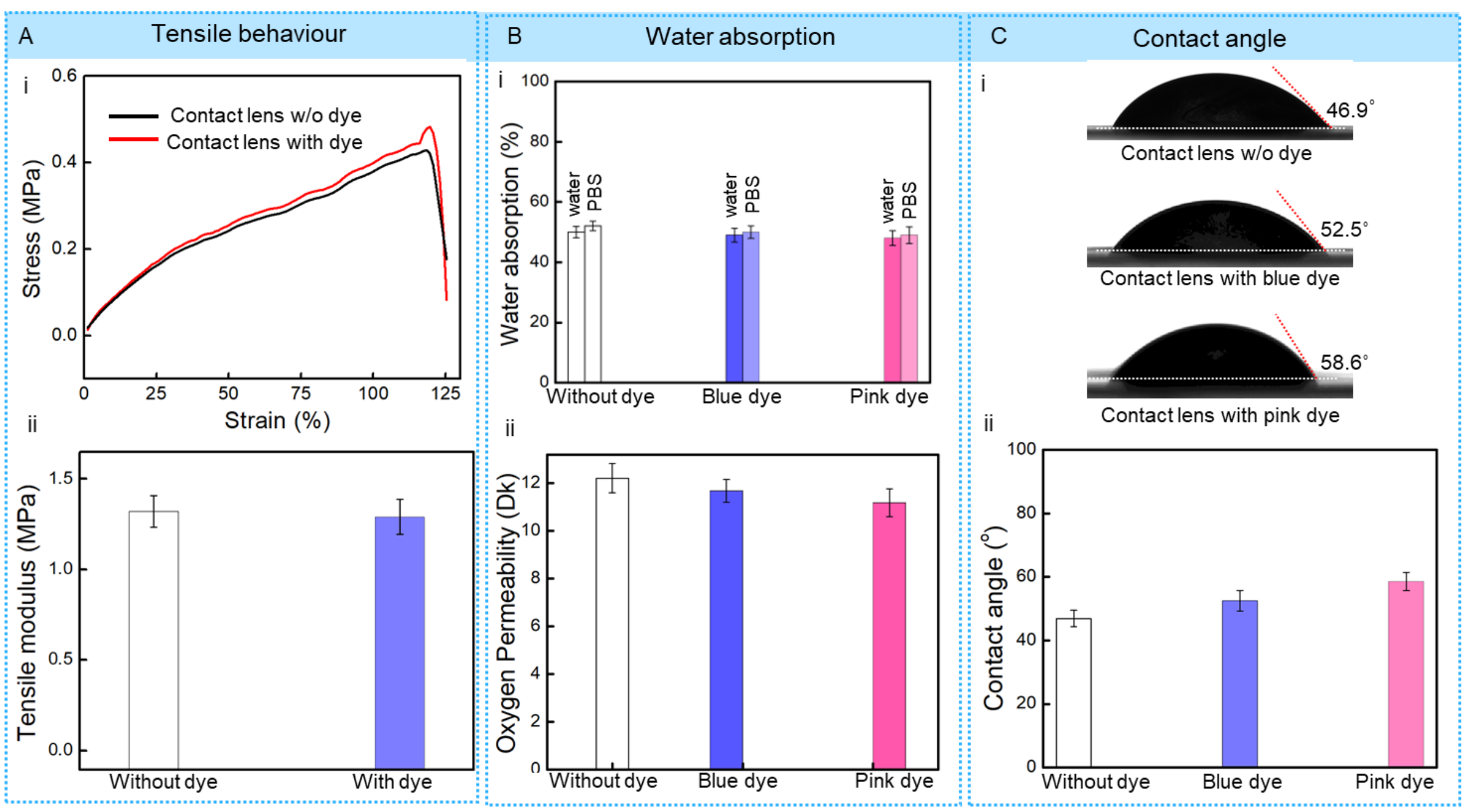

**Fig. 4**. Mechanical, swelling, and wettability characteristics of photochromic contact lenses. (A) Tensile behaviour, (i) Stress–strain curves of lenses with and without photochromic dye show comparable elasticity and fracture behaviour. (ii) Corresponding tensile modulus values indicate only a minor increase upon dye incorporation. (B) Water absorption, (i) Water uptake kinetics in deionized water and PBS. (ii) Experimentally calculated oxygen permeability (Dk). (C) Contact angle, (i) representative droplet images of contact angles for lenses without and with embedded dye. (ii) Quantified contact angles confirm a modest reduction in surface hydrophilicity after dye incorporation. The error bars represent standard deviations (SDs) between five different lenses.

### *2.3. Light-adaptable pupil-aperture scalability of the artificial iris*

Figure 5A illustrates the custom optical measurement system used to quantify wavelength-dependent transmission changes during photochromic activation. A collimated white-light source was directed through the contact lens sample, while a UV light source irradiated the lens from an orthogonal direction to initiate the photochromic reaction. The transmitted white light was collected by a spectrometer, enabling continuous acquisition of the full transmission spectrum (300–700 nm). This configuration isolates the optical effect of photochromic switching by allowing UV-induced coloration while transmission is measured independently, thereby minimizing artifacts from scattered UV light or excitation stray light.

Figure 5B compares the intrinsic optical transmission characteristics of a clear contact lens and lenses containing photochromic dyes in their unactuated (colorless) state. All samples exhibited high transparency in the visible region, with transmittance above ~82% from 450 to 700 nm, confirming that dye incorporation at the selected concentration did not significantly compromise lens clarity. In addition, all three samples (clear, UV-Pink, and UV-Blue) showed strong UV attenuation below 340 nm, where transmittance approached zero, reflecting the intrinsic UVB-blocking capability of the HEMA: PEGDA hydrogel. The UV-Pink and UV-Blue lenses further introduced additional absorption in the UVA region, resulting in a pronounced transmission decrease between approximately 360 and 450 nm.

Figure 5C and Figure S2A demonstrate the dynamic and dose-dependent coloration behavior of lenses containing the blue-switching photochromic dye. Successive UV exposure (0–20 s) produced progressive decreases in transmission across a broad spectral range of approximately 450–700 nm. The strongest attenuation occurred near 600 nm, where transmittance decreased from ~92% to ~47% (≈45 percentage-point reduction). At 400 nm, transmission decreased from ~67% to ~37%, while at 375 nm it decreased from ~46% to ~24%. This spectral modulation is consistent with reversible formation of the colored closed-ring isomer. The reduced transmission plateau at longer exposure times indicates saturation of the photostationary state. Increased absorbance in the UV region is advantageous for suppressing harmful radiation and protecting ocular tissues. Collectively, these results confirm that the blue dye exhibits efficient photoactivation and broad optical attenuation suitable for mimicking iris-induced light suppression.

Figure 5D and Figure S2B provide analogous results for lenses containing pink-switching dye. UV activation induced a reduction in visible-light transmission with a principal attenuation band centered between approximately 430 and 600 nm, consistent with the pink coloration pathway. The largest decrease was observed at 538 nm, where transmittance declined from ~89% to ~51%. At 400 nm, transmission changed from ~59% to ~39%, while at 375 nm it decreased from ~34% to ~23%. Compared with the blue dye, the overall attenuation was slightly lower, consistent with lower effective absorptivity at equivalent loading. The steady increase in UV absorption during irradiation confirms efficient photochemical switching. Figure S4 shows the dynamic optical response of the photochromic contact lens under different UV light intensities. The lens exhibited a clear intensity-dependent coloration behavior, where increasing irradiation from 0.5 to 1.0 and 1.3 mW cm$^{-2}$ progressively enhanced the blue coloration and reduced optical transmission. At low intensity (0.5 mW cm$^{-2}$), only weak coloration was observed, corresponding to a high-transmission state, whereas stronger irradiation produced a darker state with greater light attenuation. This continuous rather than binary response indicates that the lens can dynamically regulate incoming light according to environmental brightness. Such behavior mimics the natural iris, which gradually adjusts pupil aperture under changing illumination conditions. These results demonstrate that dye selection enables tunable spectral attenuation, allowing customization of artificial iris performance for different optical requirements. The proposed contact lens is designed to mimic the adaptive optical function of the human iris rather than its complete biological response pathway. In the natural eye, the iris regulates retinal light exposure through neuromuscular contraction and dilation in response to visible-light intensity, whereas the present device achieves reversible light attenuation through UV-triggered photochromic switching. Although the activation mechanism differs, UV intensity is generally correlated with bright outdoor environments, enabling autonomous darkening under sunlight and recovery to a transparent state under indoor or low-UV conditions. This passive strategy provides dynamic brightness regulation together with additional UV protection, without requiring electronics or mechanical actuation. However, because the response is UV-driven rather than directly controlled by visible luminance, the current system represents a functional approximation of iris behavior and may be further improved using visible-light-responsive photochromic materials in future designs.

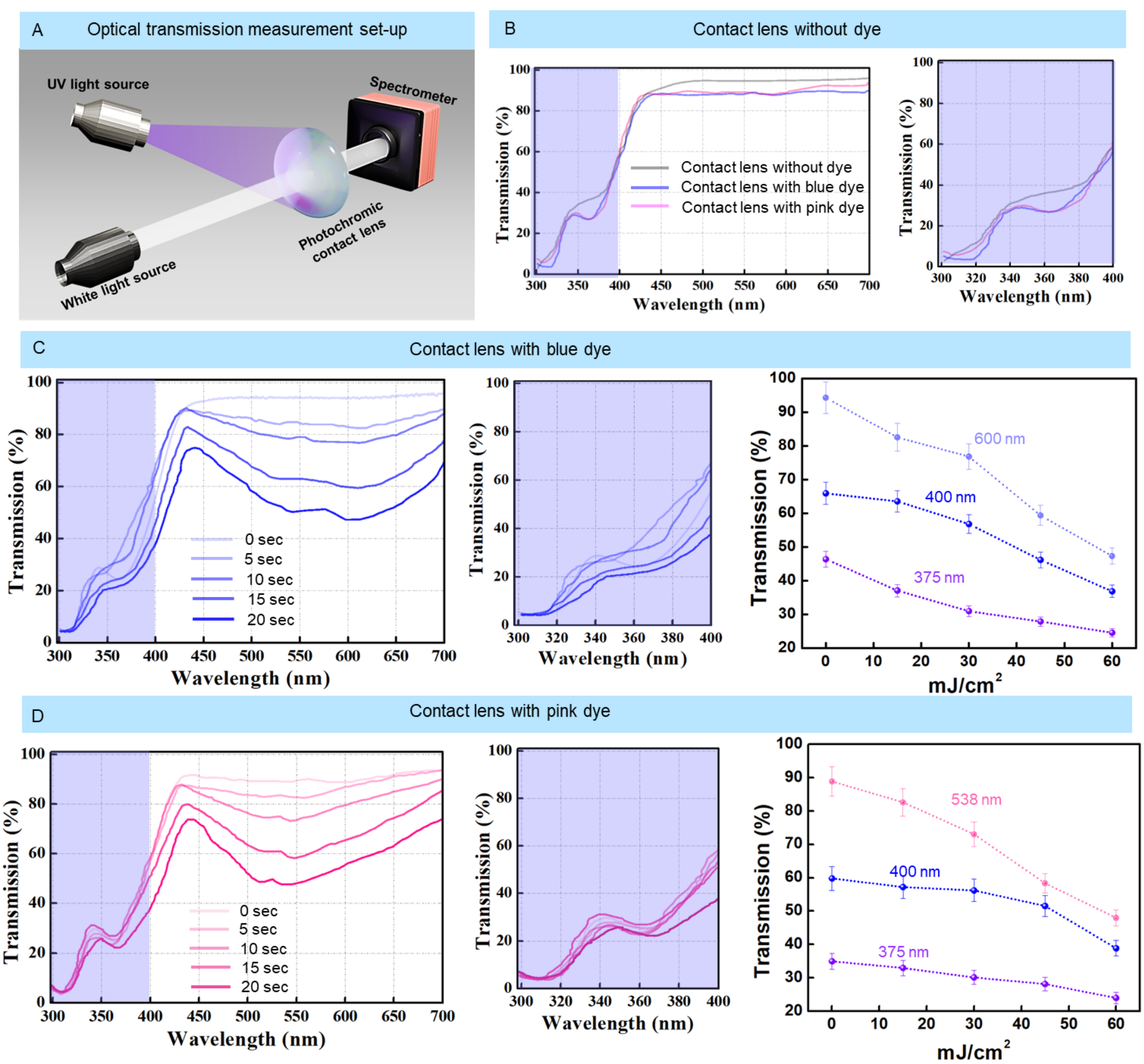

**Fig. 5.** Optical transmission characteristics of photochromic contact lenses under UV activation. (A) Schematic of the optical transmission measurement setup, showing simultaneous UV irradiation and white-light probing of the photochromic contact lens. (B) transmission spectra of lenses without dye, with blue and pink dye, highlighting comparable transparency in the visible range prior to UV activation. (C) transmission behaviour of lenses containing blue dye under UV exposure. Left: full spectra showing progressive darkening with increasing UV irradiation time

(0–20 s). Middle: UV-range spectra illustrating enhanced absorbance between 320–400 nm. Right: dose-dependent decrease in transmission at selected wavelengths (375, 400, 600 nm). (D) Transmission behaviour of lenses containing pink dye under UV exposure. Left: full spectra demonstrating photochromic darkening with increasing exposure time. Middle: UV-range transmission changes. Right: dose-dependent modulation at representative wavelengths (375, 400, 538 nm). The error bars represent standard deviations (SDs) between six different lenses.

### *2.4. Switching speed and reliability*

Figure 6 illustrates the spatial and temporal photochromic response of the contact lenses under UV activation, revealing position-dependent switching characteristics of the dynamic functional artificial iris. The schematic in Figure 6A marks four measurement locations (P1–P4) across the lens surface, progressing radially outward. The optical set-up for pointwise measurements is shown in Figure S3. For the blue-dye lens (Figure 6A), the positional transmission curves show that regions closer to the central zone (P1) undergo a weaker and slower reduction in transmission upon UV exposure, whereas outer regions (Points 2–4) exhibit progressively stronger and faster coloration, confirming a radially graded photoresponse. The corresponding bar plot (Figure 6B) quantifies the time required to reach minimum transmission at each point, demonstrating systematic increases from the inner to the outer positions, consistent with a designed gradient in dye concentration or switching kinetics. The pink-dye lens (Figure 6C, D) shows similar spatial trends but with reduced switching magnitude, reflecting the dye's lower absorbance in the measured wavelengths. Again, the time-to-minimum-transmission plot indicates slower activation compared to the blue–dye system but maintains the same radial hierarchy. Figure 6E presents repeated on/off cycling of the blue–dye lens, showing highly stable and reversible transmission changes over extended periods (>100 times) without photo-fatigue. Figure S5 presents the long-term storage stability and hydration-assisted bleaching behavior of the blue photochromic contact lens. In the dry, non-hydrated state (Day 1), the lens retained a visible blue coloration after outdoor sunlight exposure, indicating successful photoactivation and stable dye incorporation during storage. Upon hydration, the lens progressively returned toward a lighter, less colored state, demonstrating efficient bleaching and reversible photochromic recovery in an aqueous environment. After five days of continuous hydration, the lens maintained its structural integrity and reversible optical response without observable dye leakage or mechanical degradation. These results confirm favorable shelf stability together with sustained switching performance under conditions relevant to wearable contact lens use. The photochromic contact lens shows no signs of significant leakage. Figure S6 shows the leakage stability of the photochromic contact lens after soaking in DI water. The UV–VIS transmission spectra before and after soaking remain nearly unchanged, indicating preservation of optical transparency and photochromic functionality. Likewise, the overlapping FTIR spectra confirms that the hydrogel chemical structure remains intact without detectable component leaching or matrix degradation, demonstrating good hydration stability for wearable use. Figure S7 shows the effect of temperature on the photochromic switching dynamics of the contact lens by comparing responses at 25 °C and 37 °C. The lens exhibited reversible optical modulation at both temperatures, confirming stable photochromic behavior under ambient and physiological conditions. At 37 °C, the coloration process proceeded more rapidly and reached a lower minimum transmittance (~45%) than at 25 °C (~66%), indicating enhanced photoactivation at elevated temperature. Following removal of UV irradiation, both samples gradually recovered to high-transmission states (>95%) within approximately 50 s, demonstrating efficient bleaching. These results suggest that physiological temperature does not impair lens performance and may promote faster switching kinetics through increased polymer-chain mobility and dye isomerization dynamics.

Together, these results demonstrate that graded photochromic patterning enables controllable, iris-like modulation of light transmission across the lens and that the system maintains robust long-term switching performance suitable for practical artificial iris applications.

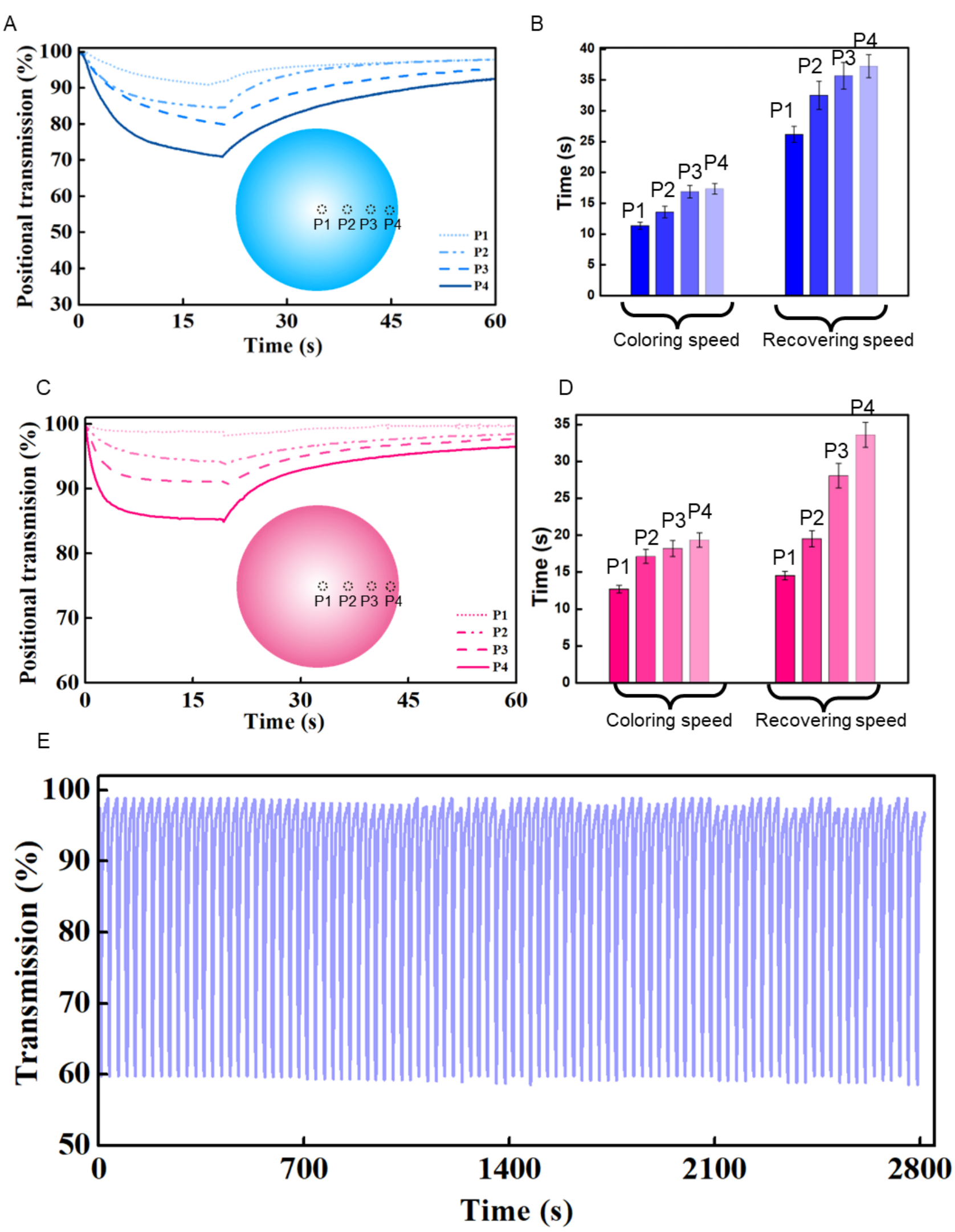

**Fig. 6.** Spatially measured photochromic response and cycling stability of photochromic contact lenses **(A)** schematic showing four measurement positions (1–4) across the contact lens surface. (B) Time-dependent transmission changes at points 1–4 for the blue-dye lens, demonstrating position-dependent darkening and recovery under UV exposure. (C) UV-irradiation time required to reach minimum transmission at each point. (D) Time-dependent transmission changes for the pink-dye lens at positions 1–4, showing similarly uniform but slightly slower kinetics compared to the blue-dye contact lens. (E) UV-irradiation time needed to reach minimum transmission for the pink-dye lens across positions 1–4. The error bars represent standard deviations (SDs) between five different lenses. (F) Repetitive UV on/off cycling of the blue-dye lens over extended periods, demonstrating stable, reversible modulation with negligible fatigue.

## 3. Conclusion

In Conclusion, we demonstrate a photochromic hydrogel contact lens that reproduces key light-adaptive functions of the human iris through spatially programmable, reversible modulation of optical transmission. The lenses retain mechanical integrity, swelling behavior, and wettability compatible with ocular use, while exhibiting rapid, wavelength-selective, and fully reversible photo-switching across the UV and visible spectrum. Engineered radial dye gradients enable precise, iris-like control of effective aperture size, and extended cycling confirms excellent photostability and repeatability. Digital micromirror–based photopatterning affords versatile spatial programmability of the hydrogel network, establishing a scalable platform for soft, power-free, light-adaptive photonic devices. This approach lays the foundation for next-generation artificial irises and ocular interfaces that combine biocompatible materials, molecular photochromism, and design-driven optical control. Looking forward, integration of photochromic hydrogel optics with adjacent photonic platforms — including whispering-gallery-mode microlasers [52], chiral-medium phase singularities [53], nonlinear thin-film photonics [54], and metasurface–quantum-photonic hybrids [55] — could further extend the functional range of soft ocular devices toward sensing, imaging, and adaptive spectral control.

## 4. Experimental section

### *4.1. Materials*

2-Hydroxyethyl methacrylate (HEMA), Poly(ethylene glycol) diacrylate (PEGDA), and Diphenyl(2,4,6-trimethylbenzoyl)phosphine oxide (TPO) were purchased from Sigma-Aldrich (USA). Digital Micromirror Device (DLP7000UV) was purchased from Texas Instruments, USA.

### *4.2. Device fabrication*

The solution used for fabrication was HEMA: PEGDA with a mixing ratio of 35:1 (in grams), with a weight ratio of 1 wt.% TPO and 2 wt. % photochromic dye. These materials were mixed for 1 hour using a magnetic stir bar set to 300 RPM at 40°C. After the solution was pipetted into a contact female mold, the male mold was placed on top of the filled female mold. It was then placed onto a sample holder and exposed to patterned UV light at 365 nm using a UV lamp (Omnicure series-2000) and a DMD for 6 minutes. A high-performance monochromatic camera was used to determine the focal plane, allowing for accurate focusing of the curing positions. The samples were then extracted and left to dry for 30 minutes. Lastly, the photochromic contact lens was peeled up from the mold.

Components of optical setup: A high-performance monochrome camera with a sensing area of 14.13 x 10.35 mm. Two lenses were used as a beam expander, with focal points of 5 and 2 cm. And one mirror is placed at 45° to reflect the UV light onto the sample mold.

### *4.3. Measurements and characterization*

Transmission spectra of the photochromic dyes were recorded on a Nicolet iS50 Fourier-transform infrared (FTIR) spectrometer. Static water contact angles of dyed and undyed hydrogel contact lenses were measured with a Rame-Hart Model 200 goniometer to assess surface wettability. Surface morphology of both the dye powders and the contact lens was examined by scanning electron microscopy (Zeiss Sigma 500). To visualize crosslinking, contact lens samples were flash-frozen in a dry ice bath and lyophilized to remove residual aqueous media (LabConco FreeZone 2.5), then sputter-coated with ~3 nm of iridium to mitigate charging and enhance image quality prior to SEM imaging. Strip-shaped specimens (5 cm length, 1 cm width, and ~700 µm thickness) were prepared to assess mechanical properties. The tensile stress tests were performed on a TA Instruments Discovery HR-30 hybrid rheometer at a constant extension rate of 100 µm $s^{-1}$. The positional dynamic response time of the artificial iris was measured with a probing He–Ne laser beam (633 nm). The transmittance curve was measured with a photodetector (Thorlabs, Inc.). The average transmittance was measured using a white-light light-emitting diode (LED) source as the probing beam. The sample was actuated by a UV LED source. The wavelength-dependent spectral transmittance was measured with an Ocean Optics spectroscope (USB2000+). The equilibrium water content (EWC = $(m\text{-}m_0)/m$) of the SP-doped hydrogel was evaluated by measuring the weight changes of the artificial iris samples in the dry case ($m_0$) and in the hydrated case ($m$) with an analytical microscale balance (Mettler-Toledo, ltd). In hydrogel materials, oxygen transport occurs primarily through the absorbed water phase. Therefore, the oxygen permeability (Dk) of hydrogels is strongly correlated with their equilibrium water uptake and can be expressed as: $Dk = 1.67e^{0.0397x Water\ absorption\ (\%)}$

**Declaration of Competing Interest**

The authors declare that they have no known competing financial interests or personal relationships that could have appeared to influence the work reported in this paper.

**Acknowledgments**

The authors would like to acknowledge financial support from 2024 Alfred P. Sloan Research Fellow, 2023 Beckman Young Investigator Award, from the Arnold and Mabel Beckman

Foundation, the Moore Foundation to the PAIR-UP Imaging Science Program, Air Force Office of Scientific Research MURI (Award No. FA9550-22-1-0312), Silicon Valley Community Foundation (Grant No. DAF2023-331948); cZi Dynamic imaging via the Chan Zuckerberg Donor Advised Fund (DAF) through the Silicon Valley Community Foundation. The work was performed in part at the San Diego Nanotechnology Infrastructure. We thank Photometrics, Inc. (Chico, CA; dissolved 2022) for support made possible through the donation honoring the research legacy of co-founder Brian Pierce (1957–2018).

**Data availability**

Data will be made available on request.

**Appendix A. Supplementary information**

Supplementary data associated with this article can be found in the online version **at doi :**

# Supplementary Information

# Digitally Programmable Photochromic Hydrogel Contact Lenses as Light-Adaptive Artificial Irises

Asad Nauman†, Tristen Moncada, Guli Gulinihali, Muhammad Waleed Khalid, Yeshaiahu Fainman and Abdoulaye Ndao*

[1] *Department of Electrical and Computer Engineering, University of California, San Diego, La Jolla, CA 92093, USA*

* Abdoulaye Ndao: a1ndao@ucsd.edu

* Corresponding author. Tel.: +1(858)-534-2721.

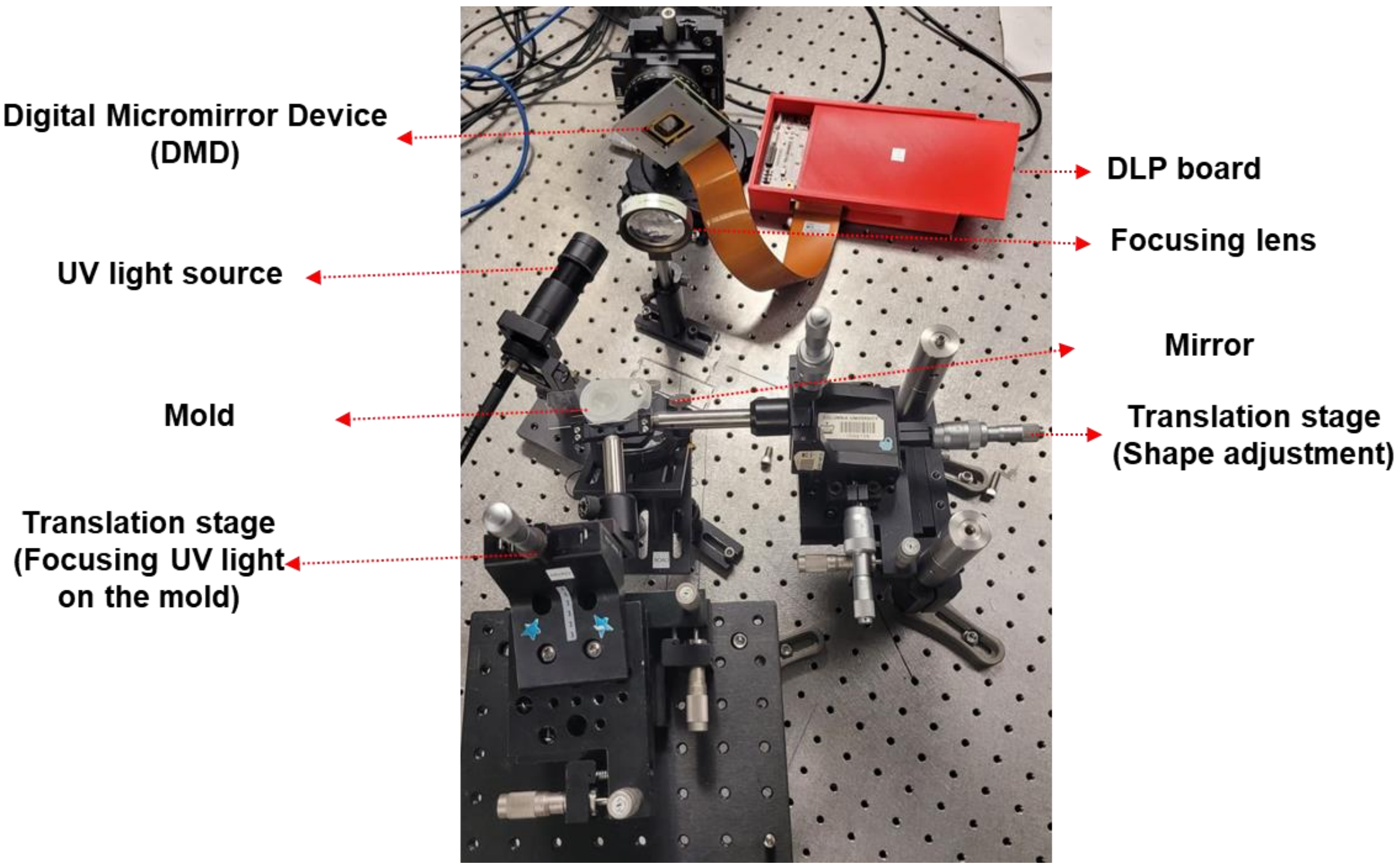

**Fig. S1.** Digital light processing set-up to pattern the UV light using DMD, where a UV light source is illuminated on the DMD and consequently shaped based on the grayscale level uploaded, which the passes through the focusing lens and the mirror to adjust the shape and the diameter of the UV light to cure the photochromic solution inside the transparent mold.

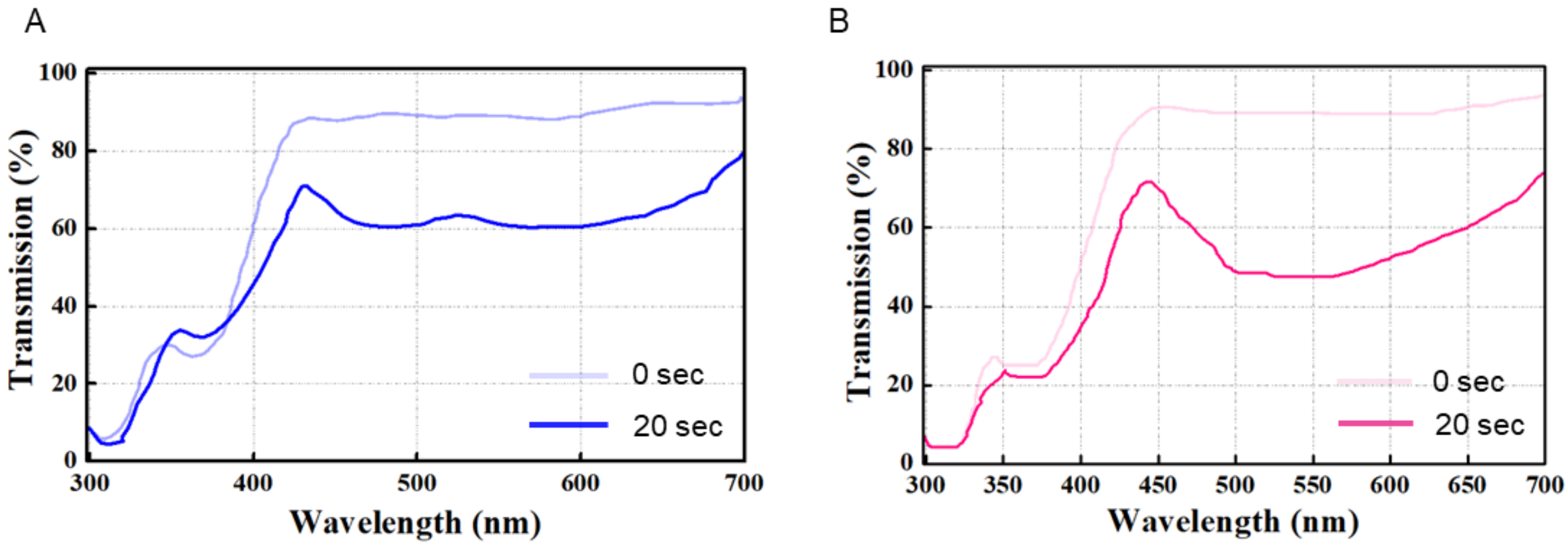

**Fig. S2.** Optical transmission characteristics of patterned photochromic contact lenses under UV activation, (A) contact lens with blue dye, (B) contact lens with pink dye.

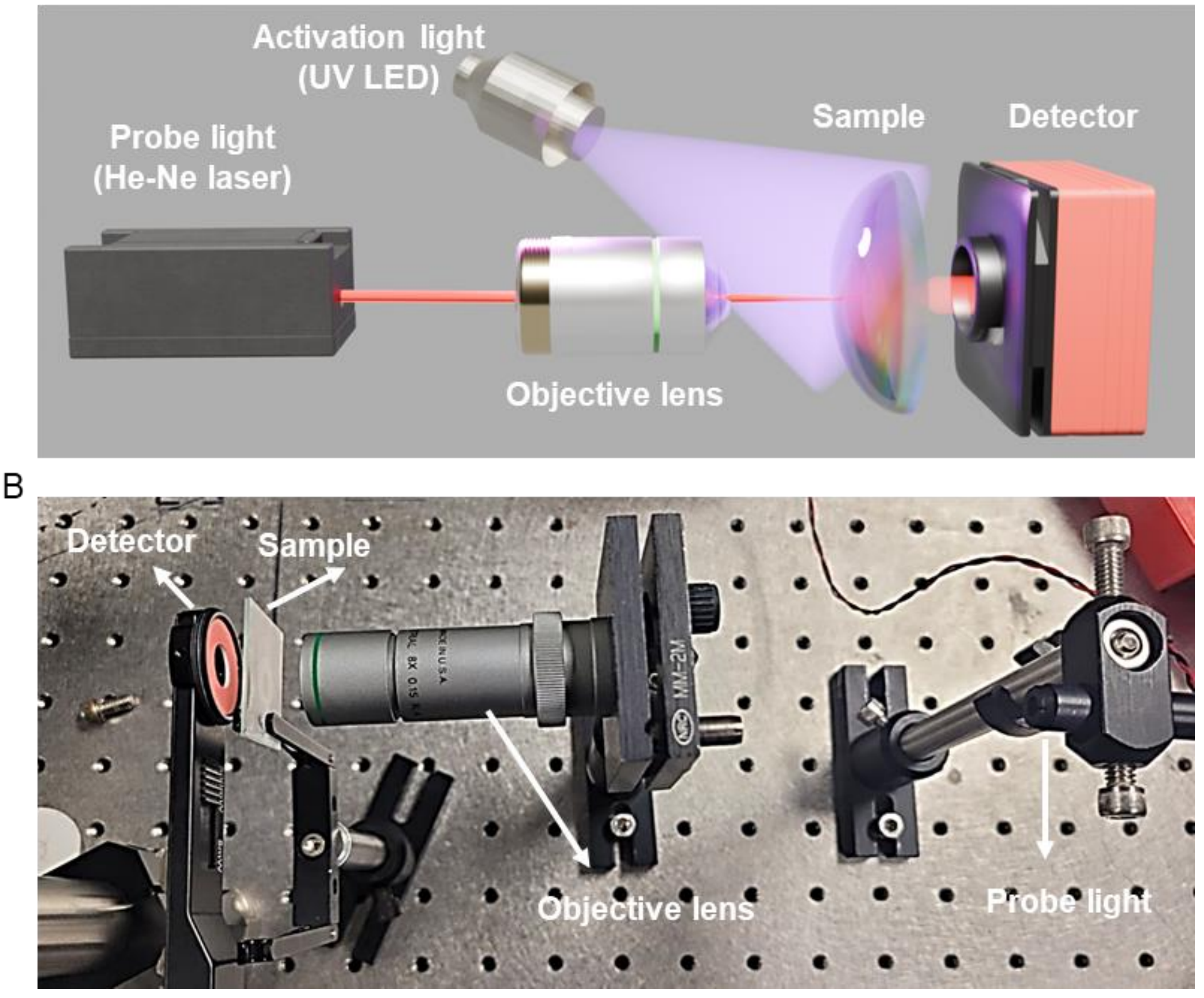

**Fig. S3.** Point-wise response-time measurement set-up of the photochromic contact lens. (A) schematic diagram, (B) experimental set-up showing helium-neon laser light work as a probe

which passes through the objective lens to control the spot size on the contact lens sample, which is then detected by the photodiode to get the intensity level as the UV light is illuminated on the contact lens sample.

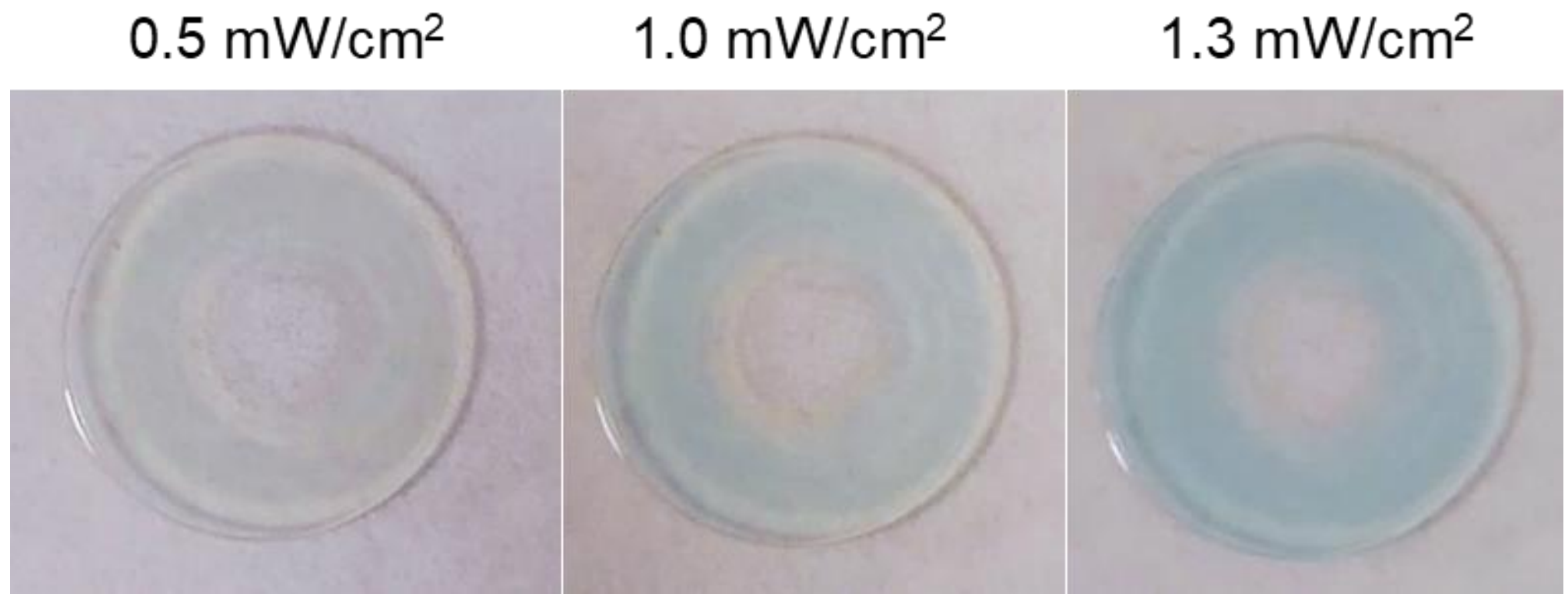

**Fig. S4.** Dynamic response of color with different UV light intensities

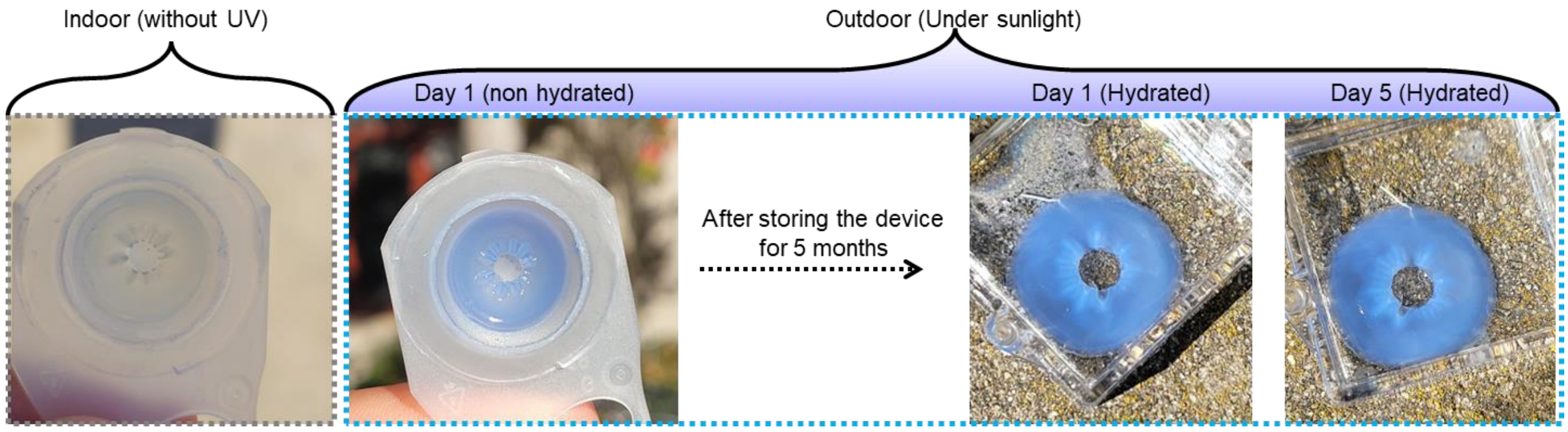

**Fig. S5.** Long term stability and bleaching effect in tear solution of the photochromic contact lens

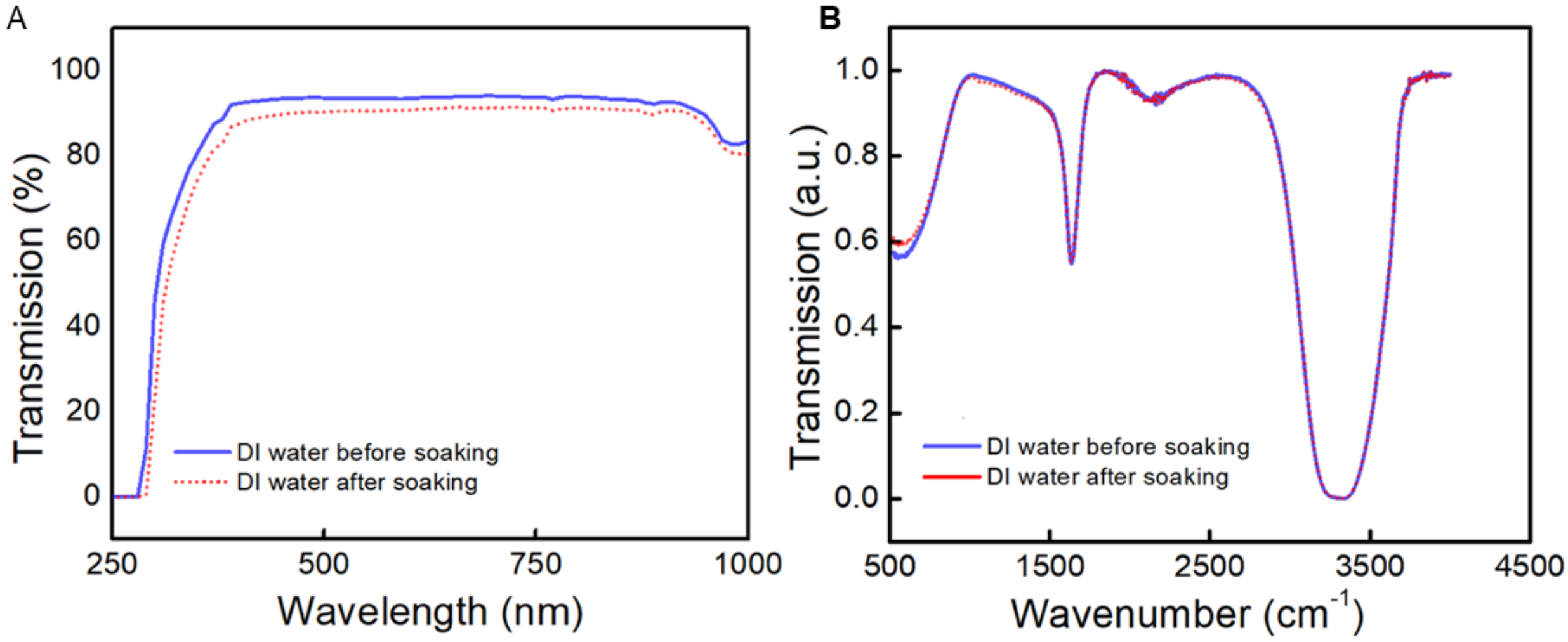

**Fig. S6.** (A) UV-VIS spectra and (B) FTIR Leakage test of the photochromic contact lens after hydration for wearable comfort.

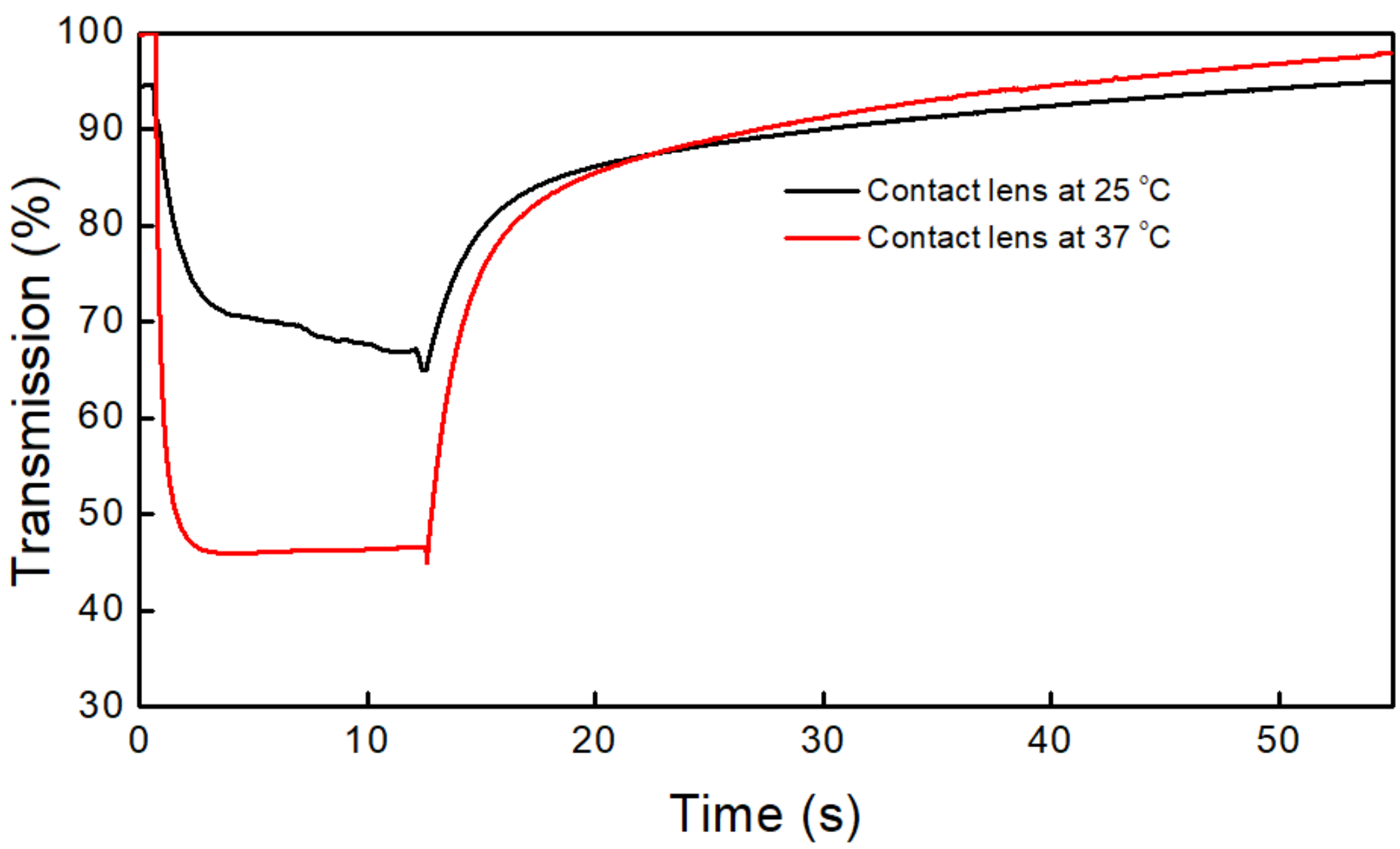

**Fig. S7.** The effect of temperature on the dynamics of the photochromic contact lens